\documentclass[useAMS,usenatbib]{mn2e}
\usepackage{epsfig,amssymb,amsmath,graphicx}

\title[Spectral and Timing Properties of 4U 1705-44]{Spectral and Timing Properties of Atoll 
Source 4U 1705-44 : {\it LAXPC/AstroSat} Results}
\author[V. K. Agrawal ]{V. K. Agrawal$\thanks{E-mail:
vivekag@isac.gov.in}$, Anuj Nandi, V. Girish and M. C. Ramadevi\\
Space Astronomy Group, ISITE Campus, ISRO Satellite Centre, Bangalore 560037, India\\
}
\begin{document}

%\date{Accepted 1988 December 15. Received 1988 December 14; in original form 1988 October 11}

\pagerange{\pageref{firstpage}--\pageref{lastpage}} \pubyear{2002}

\maketitle

\label{firstpage}

\begin{abstract}
In this paper, we present the first results of spectral and timing
properties of the atoll source 4U 1705-44 using $\sim$ 100 ks data
obtained with Large Area X-ray Proportional Counter (LAXPC) onboard {\it
AstroSat}. The source was in the high-soft state during our observations
and traced out a {\it banana track} in the Hardness Intensity Diagram
(HID). We study {\bf the} evolution of the Power Density Spectra (PDS) and the
energy spectra along the HID. PDS show  presence of a broad Lorentzian
feature (Peaked Noise or PN) centered at $1-13$ Hz and a very low
frequency noise (VLFN). The energy spectra can be described by sum of
a thermal Comptonized component, a power-law and a broad iron line. The
hard tail seen in the energy spectra is variable and contribute $4-30$\% of
the total flux. The iron line seen in this source is broad (FWHM $\sim$
2 keV) and strong (EW $\sim$ $369-512$ eV). Only relativistic smearing
in the accretion disc can not explain the origin of this feature and
requires other mechanism such as broadening by Comptonization process in
the external part of the `Comptonized Corona'. A subtle and systematic
evolution  of the spectral parameters (optical depth, electron temperature etc.)
is seen as the source moves along the HID. We study
the correlation between frequency of the PN and the spectral parameters.
PN frequency seems to be correlated with the strength
of the corona. We discuss the implication of the results in
the paper.
\end{abstract}

\begin{keywords}
accretion, accretion discs - X-rays: binaries - X-rays: individual: 4U 1705-44
\end{keywords}

\section{Introduction}
Luminous low mass X-ray binaries (LMXBs) containing weakly magnetized
Neutron Stars (NSs) are divided into two groups, called Z and atoll sources,
based on the pattern they trace out in the color-color diagram (CCD)
and the hardness-intensity diagram (HID) \citep{Hasvan89}. Z-sources
trace out a `Z' shaped path in the CCD and HID. Atoll sources trace out
a C-type curve in the CCD and HID. The discovery of the 
transient NS LMXB XTE J1701-462 \citep{Remillard06} provided a connection between
two subgroups of Z-sources (Sco-like and Cyg-like; see \citealt{Kuulkers94})
and atoll sources. At the initial phase of the outburst (at the highest
accretion rate), the source displayed behaviour like  Cyg-like Z-sources
and then as the source intensity decreased the source transformed into
Sco-like Z-sources \citep{Homan07}. At the lowest intensity state, the source
showed behaviour like atoll sources \citep{Lin09}. The observed transitions
of the source from one type to another type suggest that the accretion rate is an important 
regulating parameter which decides the behaviour of a NS LMXB on the CCD and HID.

The C-type track of atoll sources has two branches, {\it island
state} and {\it banana state}.  Luminosity and X-ray spectra of atoll
sources vary as they move from the {\it island} to {\it banana}
state. Luminosity of atoll sources varies between $0.001-0.5$ $L_{Edd}$
\citep{Done07} for a NS of mass 1.4 $M_\odot$. The {\it island
state} is characterized by low count rates and hard energy spectra. On
{\bf the} other hand, in the {\it banana state}, they have higher luminosities and
softer X-ray spectra \citep{Barret01}. The energy spectra in both states
are complex and have multiple components. Various approaches exist in
literature to model the X-ray spectra of both atoll and Z-sources. X-ray
spectra of these sources have two main components, one arising from
the accretion disc and other from the boundary layer/surface of the
NS. In one of the approaches, the spectrum is sum of the soft emission
from a standard cold disk described by multi-color-disc blackbody (MCD)
and a hard Comptonized component from the boundary layer \citep{Mit84,
Barret00, Barret01, disalvo02, agrawal03, Tarana08, agrawal09}. In an
another approach, the hot inner disc radiates a hard Comptonized component
and the boundary layer radiates a soft single temperature blackbody (BB)
\citep{disalvo00, disalvo01, Sleator16}. Sometime two thermal components,
one from the accretion disc (MCD) and another from a boundary layer (BB)
are also used \citep{Lin07, Lin10} to model the energy spectrum.

The source 4U 1705-44 is a type-I X-ray burster \citep{Got89} and is
classified as an atoll source \citep{ Hasvan89}. The source exhibited
lower kHz Quasi-periodic Oscillations (QPOs) in the frequency range
$776-866$ Hz during the RXTE (Rossi X-ray Timing Explorer) observations
\citep{Ford98} from 1997 February to June. They also reported the
detection of an upper kHz QPO at 1074$\pm$10 Hz. A strong band limited
noise (rms $\sim$ 20\%) in the power density spectra (PDS) has been seen
during the {\it island state} of the source \citep{Berger98}. The PDS in
the {\it banana state} of this source have shown presence of a low
frequency noise (LFN) represented by a broad Lorentzian centered at $\sim$
$35-45$ Hz \citep{olive03}. A narrow QPO at $\sim$ 170 Hz has also been
observed in the soft state \citep{olive03}.  In the {\it island state},
LFN centered at $9-12$ Hz and {\bf a} band limited noise with rms upto
10\% has also been reported by \cite{olive03}.

A detailed spectral behaviour of this source has been studied by
\citet{Lin10} and \citet{Piraino16}. A sum of blackbody, two Comptonized
components \citep{Piraino16} or a sum of two thermal components and a
power-law successfully describe the spectrum of this source in the soft
state. Both models require a broad iron line. A broad iron line with
FWHM of 1.2 keV has also been reported in this source by \citet{disalvo05}
 using the {\it Chandra} observations. A hard power-law tail has been seen
in this source during the soft state \citep{Piraino07,Lin10}.
During the {\it Suzaku} observations of the source in the hard state, 
signatures of reflection component are also seen \citep{disalvo15}.

In this paper, we present the first results obtained by analysis
of {\it LAXPC/AstroSat} data of the atoll source 4U 1705-44. Primary
motivation of this work is to investigate the evolution of both spectral and
temporal features of the source along the HID. We also look for possible
correlation between the spectral and temporal properties. In $\S$3,
we provide methods of analysis and modelling  of the temporal and
spectral data and in $\S$4, we present the results. In $\S$5,
we interpret and discuss the results and finally conclude in $\S$6.

\section{Observations}

Large Area X-ray Proportional Counter (LAXPC) onboard {\it AstroSat}
observed the source 4U 1705-44 from March 2, 2017 to March 5,
2017 using the Guaranteed Time (GT) observational phase of
{\it AstroSat}. LAXPC consists of 3 identical proportional counters
(LAXPC10, LAXPC20 and LAXPC30) and provides high temporal resolution and
moderate energy resolution data in the energy range of $3-80$ keV 
\citep{yadav16a, agr17, antia17}.  The combined effective area of the 
three units is $\sim$ 6000 cm$^2$ at 15 keV. The source was observed
for a total exposure time of $\sim$ 100 ks. The mode of operation was
event analysis mode. In this mode, each event is time tagged with an
accuracy of 10 $\mu$s.

\section{Data Analysis}

\subsection{Lightcurves and Hardness-Intensity Diagram}

First, we generate the background subtracted binned lightcurve in 
the $3.0-60.0$ keV energy band.  In Figure 1, we plot the lightcurve of
the source for $\sim$ 100 ks. The binsize used to create the lightcurve
is 256 seconds. During our observations, the source was in the high count
rate state (see Figure 1) and showed variabilities upto 70\% from the mean
count rate. The background subtracted lightcurves in the energy
bands: $3.0-18.0$ keV, $7.5-10.5$ keV and $10.5 - 18.0$ keV are used to
create the Hardness-Intensity Diagram (HID). The HID is shown in Figure
2. Hardness is defined as the ratio of count rates in the $10.5-18.0$
and $7.5-10.5$ keV energy bands.  Intensity is the total count rate in
the $3.0-18.0$ keV band.  In the HID, the source traced a curved
structure, referred as {\it banana state}. We divide the HID in 
seven parts `B1', `B2', `B3', `B4', `B5', `B6' and `B7',  which are shown
in Figure 2. Each segment is selected such that there are not much
changes in colour and intensity within the respective regions. In order
to carry out spectral and temporal evolution along the HID, we extract
the spectra and the lightcurves at these seven divisions of the HID.

\begin{figure}
\centering
\includegraphics[width=0.35\textwidth,angle=-90]{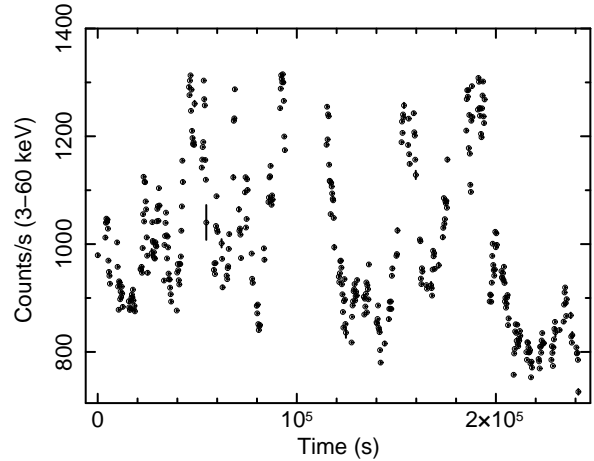}
\caption{Lightcurve of 4U 1705-44 in the $3-60$ keV energy
band. LAXPC10 data has been used to create the lightcurve. Each point
corresponds to 256 sec binsize. Error bars are smaller than the marker
size.}
\end{figure}

\subsection{Temporal Analysis} 

First, we create the lightcurves with a binsize of 0.5 millisecond in
the $3-20$ keV band. We divide the lightcurves into 16 s intervals and
create the power spectra for each of the intervals. Then we average
the power spectra belonging to the same part of the HID. We rebin the
power spectra geometrically with {\bf a} factor of 1.1 in the frequency space. These
power spectra  are normalized to give the fractional rms spectra (in the units
of (rms/mean)$^2$/Hz). Dead time corrected Poisson noise level
for a non-paralyzable detector is given by (see \citealt{Zhang95}; \citealt{yadav16b}), \begin{multline} <P(\nu)> = 2+4 \\ \times
\left(\frac{-1+\cos(2\pi \nu t_d) -(2\pi \nu \tau) \sin(2\pi \nu
t_d)} {2+(2\pi \nu \tau)^2 - 2 \cos(2\pi \nu t_d) + 2 (2\pi \nu \tau)
\sin(2\pi \nu t_d)}\right), \end{multline} where $t_d$ is dead time,
$R_T = R_{TO}/(1-R_{TO} t_d)$ is dead time corrected count rate for the
detector, where $R_{TO}$ is observed total count rate. $\tau = 1/R_T$,
where $\tau$ is average time interval between two successive events. We
apply dead time corrected Poisson noise subtraction to all the PDS
computed using this method.

It is observed that very low frequency noise (VLFN) is present in all the
seven PDS, extracted from different regions of the HID. VLFN is
modeled with a power-law ($AE^{-\Gamma}$), where $A$ is normalization and
$\Gamma$ is index of the power-law. A broad noise component (hereafter
Peaked noise component (PN)) is present in the PDS of B1-B4  and B6
parts of the HID. We model PN with a Lorentzian function given by the
following formula,
\begin{equation}
F_\nu = N/(1+(2(\nu - \nu_{PN})/FWHM)^2),
\end{equation}
where $N$ is normalization, $\nu_{PN}$ is centroid frequency and $FWHM$ is
full-width-half-maxima of the Lorentzian. In this representation, $\pi * N
* FWHM/2$ gives the integral of Lorentzian from 0 to $\infty$. We define quality factor (Q) of the PN as $\nu_{PN}/FWHM$. Errors on
the best fit parameters are computed using $\Delta \chi^2$=1.  In Figure
3, we show the PDS along with the best fit model for different parts of
the HID.  In Table 1, we show the best fit model parameters.

\begin{figure}
\centering
\includegraphics[width=0.35\textwidth,angle=-90]{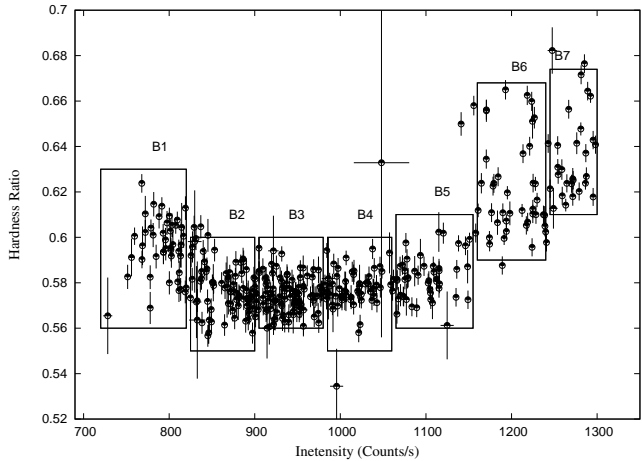}
\caption{Hardness Intensity Diagram created using LAXPC10 data. Each point corresponds to 
256 sec binsize. Different sections of HID used for spectral and temporal analysis are 
also marked. See text for details.}
\end{figure}

\subsection{Spectral Analysis}

We extract the source and background spectra for different
parts of the HID. We use the latest calibration files for the
spectral fittings. XSPEC version 12.9.1 is used to fit the spectra.
We fit the combined spectra of LAXPC10 and LAXPC20 with appropriate spectral
models. Since LAXPC30 has a different response compared to these two, we
fit it separately. We report here the results of only combined spectral
fitting of LAXPC10 and LAXPC20.

\citet{Piraino07} fitted the $0.3-200$ keV BeppoSAX spectrum with
a blackbody, a thermal Comptonization model and a broad
Gaussian line. They also detect a non-thermal power-law component
in the spectrum.  Hence, first we fit the `B1' spectrum with 
a pure thermal Comptonization model ({\it nthComp} in XSPEC; see
\citealt{Zdz96}). The reduced $\chi^2$ ($\chi^2$/dof) is 1186/158. Then,
we add a non-thermal power-law component to the model. This improved the
fit significantly ($\chi^2/dof$ = 425/156). After that we add a broad
Gaussian line component centered at $\sim$ $6.4-6.7$ keV. An addition
of this component again improved the fit significantly ($\chi^2/dof$ =
121/153). Hence, we adopt {\it nthComp+Gaussian+power-law} model to fit
all the energy spectra. The best fit parameters are shown in Table
2. We call {\it nthComp+Gauss+power-law} as {\bf Model 1}. An addition of
a blackbody component does not improve the fit.  Most probably, blackbody
component has temperature below 1 keV \citep{Piraino16} and contribution
of this component in the spectra above 3 keV is very small.  We add 1\%
systematics to overall fitting to take into account the uncertainty
in the response matrix. We also add 2\% systematics to the background
spectra to take into account the uncertainty in the background estimation.

A model consisting of emission from a multi-color disc, a blackbody
component, a power-law and a broad iron line also provides statistically
good description of the spectra  ({\bf Model 2}). This model has
been used by \cite{Lin10}. To model the blackbody, we have used {\it
bbodyrad} model in XSPEC which has normalization $N_{bb}$ = $(R_{km} /
D_{10})^2$, where $R_{km}$ is the radius of the source in km and $D_{10}$
is distance to the source in units of 10 kpc. In Figure 4, we show the
spectrum of the state `B1' fitted with {\bf Model 1} and in Figure 5,
we show the same fitted with {\bf Model 2}. Best fit model parameters
are listed in Table 2 (for {\bf Model 1}) and in Table 3 (for {\bf Model
2}). For both models, we use $N_H$ value fixed at 1.8 $\times$ 10$^{22}$
cm$^{-2}$ \citep{Lin10}. Unless quoted the errors on the best fit spectral
parameters are computed using $\Delta \chi^2$ = 1.0.

We also fit the spectra with {\it diskline + nthComp + power-law}
model. {\it Diskline} model describes the broad iron line and takes
into account the relativistic effects in the Schwarzschild metric
\citep{Fabian89}. We get a disc inclination $>$ 85$^\circ$. Fitting the
{\it diskline} model to the {\it Chandra} data of this source results in source
inclination in the range $55-84$$^\circ$ \citep{disalvo05}. Hence, we fix
the disc inclination at the highest value (84$^\circ$) while fitting
with this model. The best fit parameters for this model are listed in
Table 4. We refer to this model as {\bf Model 3}.

\begin{figure*}
\centering
\begin{tabular}{@{}cc@{}}
\includegraphics[width=0.29\textwidth,angle=-90]{b1-pds-rev.eps}&
\includegraphics[width=0.29\textwidth, angle=-90]{b2-pds-rev.eps} \\
\hspace{-.4in}
\includegraphics[width=0.29\textwidth,angle=-90]{b3-pds-rev.eps}&
\includegraphics[width=0.29\textwidth,angle=-90]{b4-pds-rev.eps} \\
\includegraphics[width=0.29\textwidth,angle=-90]{b5-pds-rev.eps}&
\includegraphics[width=0.29\textwidth,angle=-90]{b6-pds-new-rev.eps} \\
\hspace{-1in}
\end{tabular}
\end{figure*}
\begin{figure*}
\includegraphics[width=0.29\textwidth,angle=-90]{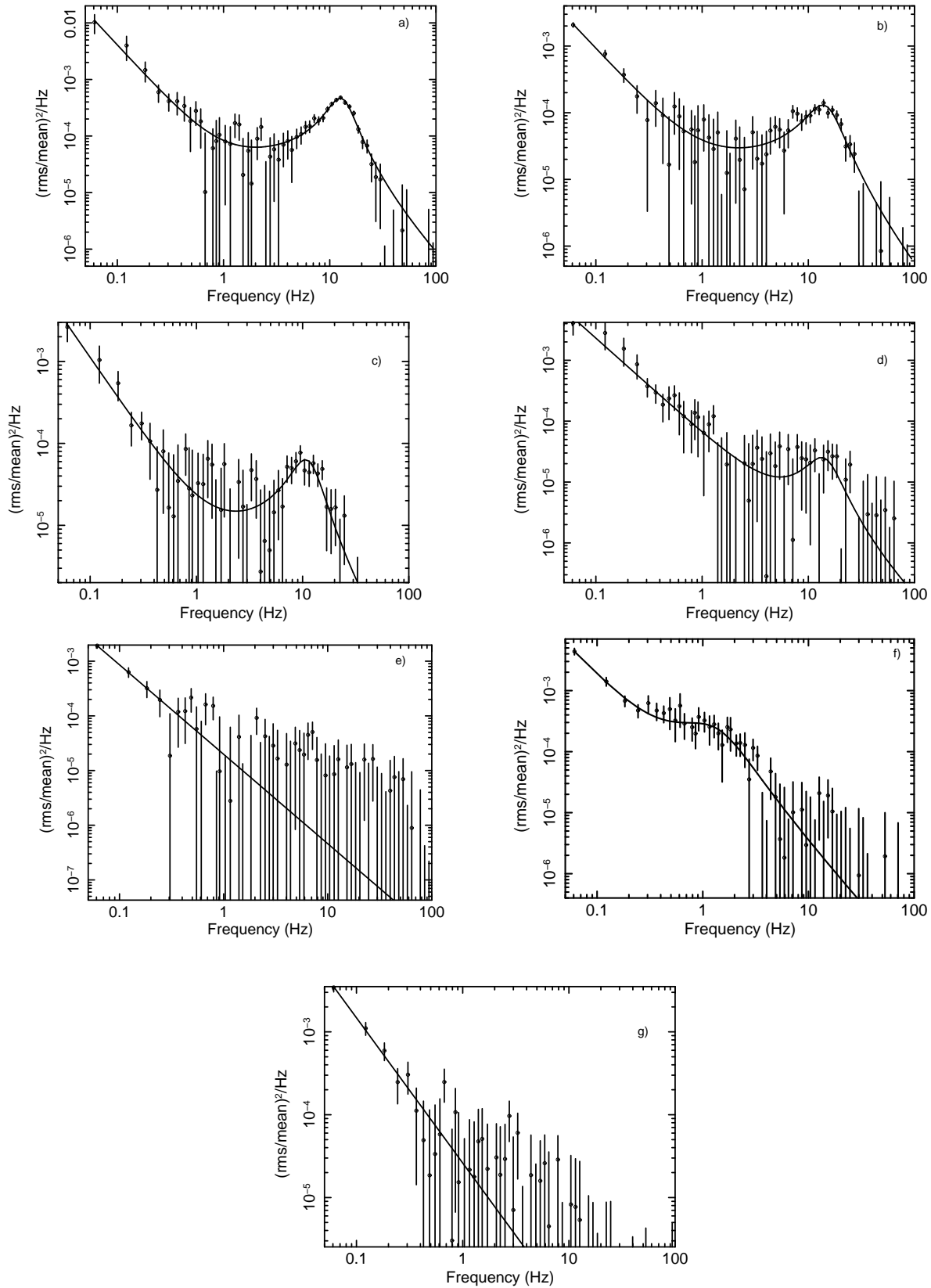}\\
\caption{The Poisson noise subtracted power density spectra of 4U 1705-44 for various phases of {\it banana state} as marked in Figure 2. Each power spectrum corresponds to different states of B1 (a), B2 (b), B3 (c), B4 (d), B5 (e), B6 (f) and B7 (g) respectively. Solid line shows the best fit model.}
\end{figure*}

\section{Results}

\subsection{Timing Behaviour}

The properties of PN and VLFN vary along the HID. The PN is centered at
$\sim$ $1-13.5$ Hz. PN frequency does not show any clear correlation
with the position on the HID. PN rms decreases from 7.1$\pm$0.23\%
to 1.87$\pm$0.61\% as the source moves from B1 to B4 (see Table 1). As
the source moves from B4 to B5, the PN component disappears and only
power-law noise remains.  At B6, it again appears with a much
lower frequency of 0.95$\pm$0.26 Hz. As the source further moves to B7,
the PN again disappears and PDS is dominated by a pure power-law
noise (see Figure 3). The index of power-law component {\bf do} not vary much
along the HID. The power-law indices lie in the range of $1.56-2.03$ (see
Table 1). The strength of VLFN is not correlated with the
position on the {\it banana branch}.

\begin{table*}
\caption{ The best fit parameters obtained by fitting the PDS corresponding
to seven segments of the HID. Power spectra of first four (B1, B2, B3,
B4) and B6 segment are fitted with a Lorentzian plus power-law model and
fifth and seventh segment (B5 and B7) are fitted with only a power-law
model. We also give percentage rms in the frequency range $0.06-100$
Hz for Lorentzian and power-law components.}

\begin{tabular}{|c|c|c|c|c|c|c}
\hline
\hline
State & \multicolumn{2}{|c|}{VLFN} & \multicolumn{3}{|c|}{Peaked-Noise (PN)} & $\chi^2/dof$\\
      & \% rms &$\Gamma$ &\% rms & $\nu_{P}$ (Hz) & FWHM (Hz) \\
\hline
B1     &2.82$\pm$0.58& 2.03$\pm$0.19 & 7.10$\pm$0.23&12.58$\pm$0.14&7.62$\pm$0.36& 70/62\\
B2     &1.44$\pm$0.25& 1.70$\pm$0.11&4.43$\pm$0.27 &13.70$\pm$0.39&11.22$\pm$0.95 &74/62\\
B3     &1.52$\pm$0.40& 1.87$\pm$0.21 &2.63$\pm$0.79&10.66$\pm$0.63&7.99$\pm$4.1 & 114/62\\
B4     &2.45$\pm$0.42& 1.56$\pm$0.13 & 1.87$\pm$0.61&13.43$\pm$2.58 &10.75$\pm$3.9 &41/62\\
B5     &1.45$\pm$0.36& 1.63$\pm$0.17 & - & - & - & 48/62\\
B6     &1.93$\pm$0.48& 1.79$\pm$0.19 &2.44$\pm$0.46&0.95$\pm$0.26 & 2.01$\pm$0.42 &80/62 \\
B7     &1.81$\pm$0.30& 1.75$\pm$0.12 & - & - & - & 82/62 \\
\hline
\end{tabular}
\end{table*}

\begin{table*}
\caption{ The best fit spectral parameters for the seven different parts
of the HID of {\bf Model 1}. $\Gamma_{nth}$ is the power-law photon index
for {\it nthComp}, $kT_e$ is the electron temperature in keV and $N_{nth}$
is the normalization of {\it nthComp}. $\Gamma_{PL}$ is the photon index
of power-law and $N_{PL}$ is the normalization of power-law. $E_{Fe}$
is the centroid energy of the iron line, $\sigma_{Fe}$ is the width of
iron line, $\tau$ is the optical depth, $y-para$ is the Comptonization
parameter. $F_{comp}$, $F_{PL}$ and $F_{tot}$ are flux of Comptonized
component, flux of power-law component and total flux respectively. Fluxes
are estimated in $0.1-100$ keV band and are in units of ergs/s/cm$^2$.
EW is the equivalent width of the iron line in eV.}
\begin{tabular}{cccccccc}
\hline
\hline
Parameters & B1 & B2 & B3 & B4 & B5 & B6 & B7\\
\hline
$\Gamma_{nth}$ & 2.10$\pm$0.01 & 2.10$\pm$0.02 & 2.05$\pm$0.04 & 1.98$\pm$0.03 & 1.96$\pm$0.02&1.88$\pm$0.02 & 1.72$\pm$0.03 \\
kT$_e$ (keV) & 2.71$\pm$0.03 & 2.69$\pm$0.03 & 2.62$\pm$0.05 & 2.57$\pm$0.04 & 2.59$\pm$0.04 & 2.57$\pm$0.04 & 2.50$\pm$0.02 \\
$N_{nth}$ & 2.78 $\pm$0.07 & 1.04$\pm$0.03 & 0.19$\pm$0.11 & 0.90$\pm$0.11 & 1.09$\pm$0.05 & 1.91$\pm$0.08 & 0.66$\pm$0.06\\
$kT_W$ (keV)   &0.18 (fixed) & 0.46$\pm$0.05 & 0.91$\pm$0.09 & 0.49$^{+0.27}_{-0.07}$ & 0.47$\pm$0.06 & 0.28 (fixed)  & 0.49$\pm$0.04 \\
$\Gamma_{PL}$ &0.86$\pm$0.21 & 0.79$\pm$0.13 & 3.41$\pm$ 0.03 & 2.51$^{+0.56}_{-0.35}$ &1.63$\pm$0.15 & 1.82$\pm$0.18 & 2.97$\pm$0.02\\
$N_{PL}$ $\times$ 10$^{2}$ &  0.17 $\pm$0.08  & 0.13 $\pm$0.06 & 746$\pm$51 & 38$^{+92}_{-23}$ & 2.11$\pm$0.8 & 5.84$\pm$1.7 & 304$\pm$62\\
$E_{Fe} (keV)$ & 6.45$\pm$0.11 & 6.51$\pm$0.13 &6.42$\pm$0.21 &6.43$\pm$0.13 & 6.44$\pm$0.14 & 6.39$\pm$0.13 & 6.41$\pm$0.09 \\
$\sigma_{Fe}$ (keV) & 1.10$\pm$0.09 & 1.16$\pm$0.06 & 1.16$\pm$0.19 & 1.14$\pm$0.07 & 1.25$\pm$0.06 & 1.26$\pm$0.11 & 1.25$\pm$0.06\\
$E_W$ (eV) & 369.5$\pm$30.8& 425.9$\pm$25.5 & 406.2$\pm$38.1 & 460.7$\pm$27.60 &511.5$\pm$20.4 & 488.7$\pm$28.6 &501.4$\pm$29.8\\ 
$\tau$        & 9.78$\pm$0.09 & 9.82$\pm$0.17 & 10.32 $\pm$0.61 & 10.94$\pm$0.34 & 11.06$\pm$0.23 & 11.78$\pm$0.27 & 13.69$\pm$0.56 \\
$y-$para & 2.03$\pm$0.03 & 2.03$\pm$0.07 & 2.18$\pm$0.25 & 2.40$\pm$0.14 & 2.48$\pm$0.10 & 2.79$\pm$0.12 & 3.67$\pm$0.31 \\
$R_{seed} (km)$ & 82.65$\pm$28.5 & 22.70$\pm$1.48 & 7.03$\pm$0.39 & 22.12$\pm$1.47 & 24.85$\pm$1.37&56.87$\pm$17.32&22.95$\pm$0.89\\
$F_{comp}$ $\times 10^{9}$& 7.29$\pm$0.31 & 7.76$\pm$0.45 & 5.62$\pm$0.26 & 8.51$\pm$0.71 & 9.54$\pm$0.25&10.45$\pm$0.50&8.91$\pm$0.42\\
$F_{PL}$ $\times 10^{9}$ & 0.467$\pm$0.05 & 0.457 $\pm$0.04 & 3.01$\pm$0.29& 0.676$\pm$0.27& 0.371$\pm$0.05 &0.586$\pm$0.07&2.55$\pm$0.12\\
$F_{tot}$ $\times 10^{9}$ &  7.97$\pm$0.35 & 8.48$\pm$0.55 & 8.89$\pm$0.39 & 9.53$\pm$0.76 &10.04$\pm$0.25 & 11.47$\pm$0.49 &11.89$\pm$0.43 \\
$\chi^2/dof$ & 121/153 & 82/153 & 91/153 & 111/153 &142/153 &120/153 & 104/153\\
\hline
\hline
\end{tabular}
\end{table*}

\begin{table*}
\caption{ The best fit spectral parameters of {\bf Model 2}. $kT_{in}$
is the inner disc temperature in keV and $N_{dbb}$ is the normalization
of diskbb component. $kT_{bb}$ is the blackbody temperature in keV and
$N_{bb}$ is the normalization of blackbody. $\Gamma_{PL}$ is the photon
power-law index and $N_{PL}$ is the normalization of power-law. $R_{in}$
and $R_{BB}$ are the inner disk radius and boundary layer radius
respectively in units of km. $R_{eff}$ is the effective radius of inner
disc in km. $F_{BB}$, $F_{dbb}$, $F_{PL}$, $F_{tot}$ are blackbody
flux, disk flux, power-law flux and total flux respectively. Fluxes are
estimated in $0.1-100$ keV band and are in units of ergs/s/cm$^2$.}
\begin{tabular}{cccccccc}
\hline
\hline
Parameters & B1 & B2 & B3 & B4 & B5 & B6 & B7\\
\hline
$kT_{in}$ {(keV)}& 1.18$\pm$0.03 & 1.28$\pm$0.04 & 1.78$\pm$0.14 & 1.68$\pm$0.08 & 1.62$\pm$0.10 & 1.56$\pm$0.11 & 1.73$\pm$0.12 \\
$N_{dbb}$& 170.5$\pm$18.5 &133.4$\pm$16.5 & 25.8$\pm$8.3  & 35.3$\pm$8.6 & 45.89 $\pm$11.6 & 51.3$\pm$14.7 & 35.83$\pm$9.82 \\
$kT_{bb}$ (keV)& 2.18$\pm$0.02 & 2.22$\pm$0.03 & 2.41$\pm$0.08 &2.33$\pm$0.05 & 2.33$\pm$0.04&2.30$\pm$0.04 & 2.38$\pm$0.05 \\
$N_{bb}$ &11.93$\pm$0.69 & 11.66$\pm$0.88 & 5.78$\pm$1.71 & 8.6$\pm$1.44 &  10.31$\pm$1.57 & 13.65$\pm$2.13 & 12.5$\pm$2.16 \\
$\Gamma_{PL}$ & 1.74$\pm$0.21 & 1.62$\pm$0.26 &3.29$\pm$0.11 &3.20$\pm$ 0.15 & 2.90$\pm$0.20&2.83$\pm$0.16 & 2.98$\pm$0.11 \\
$N_{PL}\times$ 10$^{2}$ &4.8$\pm$2.1 & 2.9$\pm$1.1 & 508 $\pm$124 & 438$\pm$152 & 188$\pm$61&216$\pm$84&325$\pm$81\\
$R_{in}$ (km)& 23.79$\pm$1.28 & 21.05$\pm$1.30 &9.26$\pm$1.48 & 10.82$\pm$1.31 & 12.24$\pm$1.57 &12.97$\pm$1.88 &10.90$\pm$1.49\\
$R_{BB}$ (km)& 2.59$\pm$0.075 & 2.56$\pm$0.09 & 1.80$\pm$0.27 & 2.19$\pm$0.18&2.40$\pm$0.18 &2.77$\pm$0.21 & 2.65$\pm$0.22\\
$E_{Fe}$ (keV) & 6.44$\pm$0.09 & 6.49$\pm$0.08 &6.38$\pm$0.22 &6.43$\pm$0.31&6.65$\pm$0.10&6.70$\pm$0.09 & 6.56$\pm$0.08\\
$\sigma_{Fe} (keV)$ & 1.30$\pm$0.06 &1.31$\pm$0.09 &1.19 $\pm$0.11 &1.16$\pm$0.21 &0.99$\pm$0.09&1.05$\pm$0.09 & 1.04$\pm$0.08\\
$R_{eff}$ (km) &68.77$\pm$3.72 & 60.83$\pm$3.76 & 26.77$\pm$4.30 & 31.27$\pm$3.84 & 35.39$\pm$4.55&37.48$\pm$5.44&31.52$\pm$4.30\\
$F_{BB}$ $\times 10^{9}$ & 2.88$\pm$0.13& 3.01$\pm$0.12 & 2.08$\pm$0.31 & 2.71$\pm$0.26 & 3.23$\pm$0.39&3.98$\pm$0.38 & 4.07$\pm$0.45\\
$F_{dbb}$ $\times 10^{9}$ & 3.89$\pm$0.19 &4.36$\pm$0.31 &4.07$\pm$0.55 & 4.16$\pm$0.64 &4.57$\pm$0.35 & 4.68$\pm$0.33 &4.89$\pm$0.35\\
$F_{PL}$ $\times 10^{9}$ & 0.631$\pm$0.09 & 0.562$\pm$0.25 & 2.45$\pm$0.32 & 2.46$\pm$0.39 & 1.65$\pm$0.51&2.18$\pm$0.21 & 2.39$\pm$0.23 \\
$F_{tot}$ $\times 10^{9}$ & 7.70$\pm$0.25 & 8.28 $\pm$0.39 & 8.90$\pm$0.88 & 9.66$\pm$ 1.15 & 9.74$\pm$0.68& 11.16$\pm$0.55&11.70$\pm$0.67 \\
$\chi^2/dof$ & 135/153 & 101/153 &93/153 & 141/153&149/153 &130/153 & 107/153\\
\hline
\hline
\end{tabular}
\end{table*}
\begin{table*}
\caption{ The best fit spectral parameters of {\bf Model 3}.}
\begin{tabular}{cccccccc}
\hline
\hline
Parameters & B1 & B2 & B3 & B4 & B5 &  B6 & B7\\
\hline
$R_{in} (GM/c^2)$ & 6.0$^{+0.33}_{-0.04}$ &  6.0$^{+1.52}_{-0.05}$ & 6.0$^{+2.64}_{-0.04}$ & 6.0$^{+1.15}_{-0.05}$ & 6.0$^{+0.65}_{-0.04}$ & 6.0$^{+0.14}_{-0.03}$ & 6.0$^{+0.18}_{-0.03}$ \\  
$E_{Fe} (keV)$ & 6.38$\pm$0.15 & 6.50$\pm$0.06 & 6.52$\pm$0.04 & 6.42$\pm$0.06 & 6.41$\pm$0.07 & 6.39$\pm$0.05 & 6.42$\pm$0.06 \\ 
$\beta$ & -2.31$^{+0.07}_{-0.20}$ &  -2.27$^{+0.08}_{-0.17}$ & -2.44$^{+0.13}_{-0.23}$ &  -2.37$^{+0.07}_{-0.16}$ & -2.43$^{+0.06}_{-0.06}$ & -2.44$^{+0.09}_{-0.06}$ &    -2.44$^{+0.05}_{-0.02}$ \\ 
$\Gamma_{nth}$ & 2.10$\pm$0.02 & 2.11$\pm$0.03 & 2.11$\pm$0.07 & 1.99$\pm$0.04 & 1.98$\pm$0.02 & 1.88$\pm$0.02 & 1.77$\pm$0.05 \\
$kT_e (keV)$ & 2.67$\pm$0.03 & 2.70$\pm$0.04 & 2.66$\pm$0.08 & 2.55$\pm$0.04 & 2.56$\pm$0.04 & 2.56$\pm$0.02 & 2.52$\pm$0.04 \\ 
$kT_W$ (keV)& 0.18 (fixed) &0.52$\pm$0.05 & 0.95$\pm$0.05 & 0.85$^{+0.09}_{-0.31}$ & 0.55$\pm$0.02 & 0.30 (fixed) &  0.86$\pm$0.15\\ 
$N_{nth}$ & 2.79$\pm$0.04 & 0.84$\pm$0.09 & 0.18$\pm$0.04 & 0.26$\pm$0.05 & 0.82$\pm$0.06 & 1.82$\pm$0.05 & 0.24$\pm$0.06 \\
$\Gamma_{PL}$ & 0.97$\pm$0.17 & 0.87$\pm$0.25 & 3.43$\pm$0.09 & 3.29$^{+0.08}_{-0.65}$ & 1.84$\pm$0.14 & 1.82$\pm$0.12 & 3.17$\pm$0.13\\
$N_{PL} \times 10^2$ & 0.27$\pm$0.09 &  0.18$\pm$0.11 & 748$\pm$128 & 567$^{+146}_{-376}$ & 4.4$\pm$0.12 & 6.14$\pm$1.9 & 603$\pm$ 176 \\
$\chi^2/dof$ & 123/152 & 83/152 & 95/152 & 113/152 & 149/152 &132/152 & 108/152 \\
\hline
\end{tabular}
\end{table*}

\subsection{Spectral Properties} 

In Figure 6, we show the variation of the best fit parameters of
{\bf Model 1}. The spectra consist of a thermal Comptonization
component ({\it nthComp}), a power-law emission and a broad Gaussian
line. The spectra of the source in the $3-80$ keV band do not
require a soft thermal component and probably observation in the
soft energy band ($<$ 3 keV) is required to observe this component. The
parameters of {\it nthComp} are the electron temperature, power-law
photon index and seed photon temperature. We assumed blackbody as the
source of the seed photon. We note that the spectra during
our observations were soft. The electron temperature $kT_e$ is in the
range of $2.5-2.71$ keV. Fitting the data with this model gives 
the seed photon temperature of $0.45-0.9$ keV for B2, B3, B4, B5 and
B7. However, the seed photon temperature for B1 and B6 can not be
constrained and we need to fix it at the best fit value of
0.18 keV and 0.28 keV respectively.

The electron temperature shows subtle variations as the source moves from
B1 to B7. The electron temperature $kT_e$ decreases from 2.71$\pm$0.03
keV to 2.50$\pm$0.02 keV as the source moves from B1 to B7. At the same
time the spectral shape becomes harder and photon index
$\Gamma_{nth}$ decreases from 2.10$\pm$0.01 to 1.72$\pm$0.03. We also
derive the optical depth using the formula
\citep{Zdz96},

\begin{equation}
\Gamma_{nth} = \left[\frac{9}{4}+\frac{m_e c^2}{kT_e\tau(1+\tau/3)}\right]^{\frac{1}{2}} - \frac{3}{2},
\end{equation}
where, $m_e$ is {\bf the} mass of the electron, $c$ is {\bf the} speed of light in vacuum, $\tau$ is the optical depth. Here, the optical depth $\tau$
approximately equals the radial optical depth for a uniform spherical
corona. The corona is found to be optically thick. The optical depth
increases from 9.78$\pm$0.09 to 13.69$\pm$0.56 as the source moves from
B1 to B7. The Comptonization parameter $y$, which is a measure of degree
of Comptonization also increases from 2.03$\pm$0.03 to 3.67$\pm$0.31
as the source moves from B1 to B7.

The power-law observed during the soft state of the source 4U
1705-44 is variable and is present in all the seven spectra taken at
different parts of the HID. The index of the power-law component varies
between $0.79-3.41$ when we model the spectra with {\bf Model 1}. The
contribution of the power-law component to the total flux 
varies from 4 to 30\% along the HID. However, it is not correlated with
the position on the HID track.

The energy spectra have degeneracy and can also be described by
{\it diskbb+BB+PL+Gauss}. In Figure 7, we plot the variation of the best fit
spectral parameters of {\bf Model 2}. The $\chi^2/dof$ is very similar for
both models. The normalization of {\it diskbb} component $N_{dbb}$ 
varies between 25 to 170. The disk radius is connected with the
normalization by formula, $N_{dbb} = (R_{in}/D_{10})^2 \cos\theta$, where
$R_{in}$ is the inner disk radius in km, $D_{10}$ is distance to
the source in {\bf 10 kpc} and $\theta$ is the disk inclination with respect
to the observer. We give the derived disk radius $R_{in}$ assuming 
a source inclination of $80^\circ$ (as indicated by {\bf Model 3}) and a distance of 7.4 kpc
\citep{Haberl95}. The disk radius is in the range of $9-23$ km.

\citet{shimura95} suggested that if electron scattering modifies the
spectrum from a  standard disc, then diluted blackbody given by
the equation,
\begin{equation}
F_\nu = \frac{\pi}{f_{col}^4} B_\nu(f_{col} T_{eff}),
\end{equation}
approximates the local spectra. Here, $f_{col}$ is the spectral
hardening or colour-correction factor and $T_{eff}$ is the effective
temperature and $B_\nu$ is the Planck function. They found that
the local spectrum can be described by a diluted blackbody
with $f_{col} \sim$ 1.7 for 10\% of Eddington luminosity. For a
fully relativistic accretion disc and taking into account the bound-free
process, \citet{davis05} derived the spectral hardening factor for
different luminosities and inclination angles. For an inclination angle
$i$ = 70$^\circ$ and $L/L_{Edd} $ = 0.3, they obtained $f_{col}$ = 1.73
and for an inclination angle $i$ = 45$^\circ$ and for same luminosity $f_{col}$
is $\sim$ 1.6. In our case, the total luminosity $L_{tot}$ is in the
range of $0.3-0.5$ $L_{Edd}$. Hence, we take spectral hardening factor
to be $\sim$ 1.7 to derive the disc radius.  The effective temperature
and effective inner disc radius is given by,
\begin{equation}
T_{eff} = T_{col}/f \quad and \quad 
R_{eff} = f^2 R_{in}.
\end{equation}
In Table 3, we also give {\it effective radius} of the inner
disc. There may be {\bf an} uncertainty in the inferred disk radius due to
uncertainty in the disk inclination and the source distance. The photon
index of power-law varies between $1.62-3.3$ when we model the
spectra with {\bf Model 2}. We also note that both flux and power-law
index do not show any clear correlation with the position on the HID
track.

\begin{figure}
\includegraphics[width=0.32\textwidth,angle=-90]{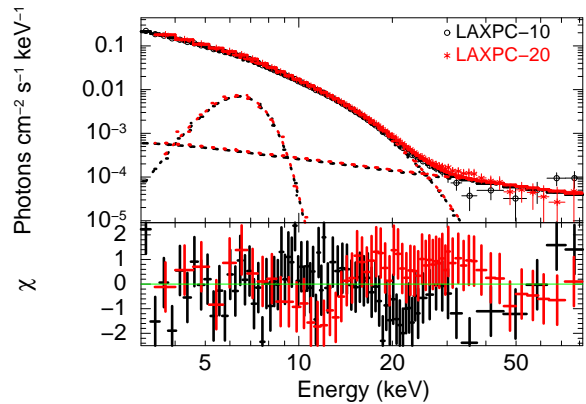}\\
\caption{Top panel shows the best fit model and unfolded spectrum. The model used is
{\it nthComp+Gauss+PL} ({\bf Model 1}). The bottom panel shows residual in units of $\sigma$.}
\end{figure}
\begin{figure}
\includegraphics[width=0.32\textwidth,angle=-90]{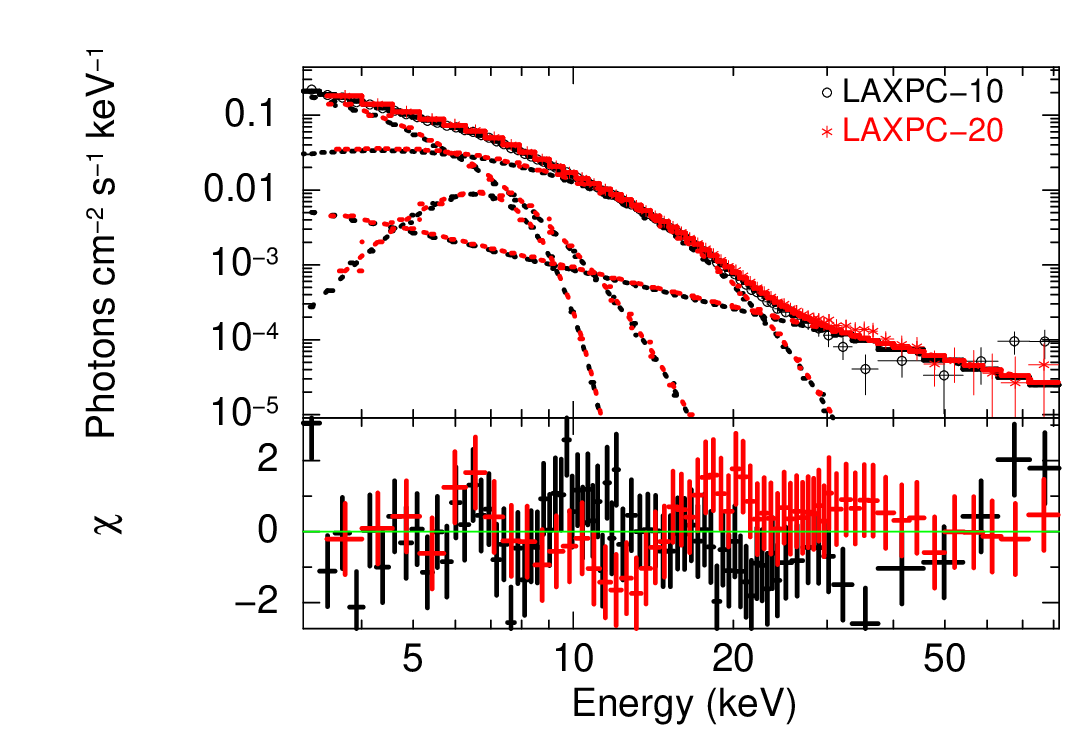}\\
\caption{Top panel shows the best fit model and unfolded spectrum. The model used is
{\it diskbb+bbodyrad+Gauss+PL} ({\bf Model 2}). The bottom panel shows residual in units 
of $\sigma$.}
\end{figure}

We find that fitting the spectra of the HID sections of B1-B7 with {\bf
Model 2} gives higher reduced $\chi^2$ compared to {\bf Model 1} (see
Table 2 and Table 3). Therefore, {\bf Model 1} is a better description
of the X-ray energy spectra of the atoll source 4U 1705-44 in the {\it
banana state} than {\bf Model 2}.

The Gaussian iron lines detected in the energy spectra are very
broad (width $\sigma_{Fe}$ $\sim$ $1.0-1.3$ keV) and also strong
(EW $\sim$ $369-$512 eV). We also note that modelling the $F_e$
line with {\it diskline} model {\bf (Model 3)} gives a disc
inclination $>$85$^\circ$, which is higher than previously reported
value \citep{disalvo05,disalvo09,Lin10,Piraino16}.  If we freeze the
inclination at 84$^\circ$ then the inner disc radius approaches
to the inner most circular orbit ($6 GM/c^2$) where, G is the
universal gravitational constant and M is mass of the central object
(see Table 4). Note that inclination angle $>84^\circ$ does not improve
the fit and a lower inclination worsen the fit. We also note that {\bf
Model 3}  which includes {\it diskline} component to model the iron line
is not a better description of the data compared to  {\bf Model 1}
(see Table 4).

\subsection{Correlation between parameters of spectra and Peaked noise}

In order to understand the possible origin of the Peaked noise
(PN), we correlate frequency of the PN with the parameters
of Comptonization. Figure 8a shows the variation of the
PN frequency vs $\tau$. There is no clear correlation between these two
parameters. In Figure 8b, we show the variation of the PN frequency vs
Comptonization flux. We find that the PN frequency is correlated
with the Comptonization flux F$_{comp}$ except the last point. In
Figure 8c, we plot the PN frequency as a function of $kT_e$. No clear
correlation is observed between these two parameters.

\section{Discussion}
We apply three different empirical models to describe the energy spectra
of the atoll source 4U 1705-44 in the soft state observed during the
GT phase of {\it AstroSat} mission. In first approach, Comptonization
model of \cite{Zdz96} describes the complex curvature in the soft
state. However, in the second approach, we fit the spectral curvature
with two thermal components: {\it diskbb} and {\it BB}. This approach has
been adopted by \cite{Lin10}. First approach does not require a thermal
component. Either the soft component may be inside the optically thick
Corona and completely Comptonized, or it is peaking at lower energies
and its contribution to the $3-80$ keV range is negligible.

\begin{figure}
\includegraphics[width=0.65\textwidth,angle=-90]{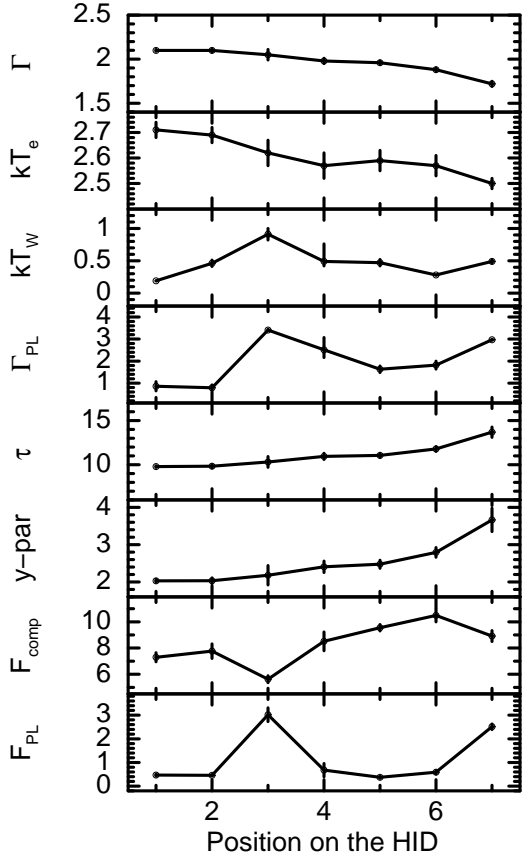}\\
\caption{ Variation of the best fit spectral parameters and model
fluxes of {\bf Model 1} along the HID. In the X-axis, 1, 2, 3, 4, 5, 6,
7 correspond to segments B1, B2, B3, B4, B5, B6 and B7 respectively. Model
fluxes are in units of 10$^{-9}$ ergs/s/cm$^2$.}
\end{figure}
\begin{figure}
\includegraphics[width=0.65\textwidth,angle=-90]{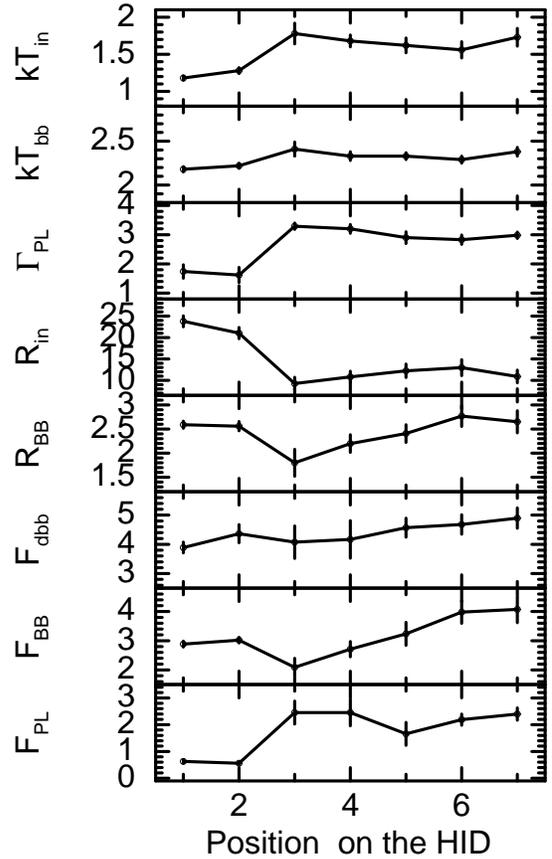}\\
\caption{ Variation of the best fit spectral parameters and fluxes of {\bf
Model 2} along the HID. In the X-axis, 1, 2, 3, 4, 5, 6, 7 correspond
to segments B1, B2, B3, B4, B5, B6, B7 respectively. Model fluxes are
in units of 10$^{-9}$ ergs/s/cm$^2$.}
\end{figure}

By fitting the spectra with the Comptonization model, we find that
the corona is cool and optically thick ($\tau \sim 10-14$). If
this corona covers the NS surface then soft BB photons emitted from
the boundary layer will be Comptonized. The enhancement of seed photon
luminosity by scattering in an optically thick Compton cloud is
given by $3kT_e/kT_{seed}$ \citep{Dermer91}. The electron temperature
of the corona is $\sim$ 3 keV. We have assumed source of 
the seed photon as a blackbody emission from the surface of NS. We
get luminosity enhancement factor $\eta \sim $ $8-42$. The seed
photon flux $F_{seed}$ will be $F_{Comp}/\eta$. In Table 2, we give the
NS radius derived from this seed photon flux assuming an isotropic
luminosity. The derived radius is in the range of $7-80$ km. Here, we
have assumed a spherical corona around the NS. However,
it could be transition layer between the NS surface and accretion
disc \citep{Titarchuk01} or Accretion Disc Corona (ADC) above the disc
\citep{church14}. Moreover, we could be able to constrain the variation in
the properties of the corona e.g. temperature and optical depth. We
find that as the source intensity increases, optical depth of the
corona also increases. Most probably radiation pressure drives matter
from the accretion disc to the larger corona or ADC \citep{agrawal03}
and hence increases its density. As the optical depth of the
corona increases, the photons suffer more number of scatterings and
hence in turn causes a decrease in the corona temperature. The
above scenario explains the decrease in the temperature of the
corona as the source moves from B1 to B7.

Fitting the data with the model adopted by \cite{Lin10} also provides
a statistically good description of the spectra. The inner-disc
temperature varies between $1.20-1.80$ keV as the source moves along
the {\it banana track}.  The effective inner disc radius $R_{eff}$
varies from $26-70$ km, assuming spectral hardening factor
$f_{col}= 1.7$.

The boundary layer emission is modeled as blackbody. The NS radius
obtained is very small $\sim$ 2 km (see Table 3). The radius of
NS is obtained assuming that the NS boundary layer emits
isotropically. However, the boundary layer may not emit
isotropically. \cite{Lin10} suggested that emission area of the
boundary layer is $\sim$ 20\% of the  NS surface area (for
inclination $i$ = 25$^\circ$). This will increase the NS radius by a
factor of 2.25 times and still {\bf gives} smaller values of NS radius ($\sim$
5 km). Even smaller size of the boundary layer depends on the
radiative transfer processes in the NS atmosphere \citep{Popham01,
Lin10} and the spectral hardening factor due to  free-free and
bound-free processes. The radius of NS/Bounadry layer ($R_{seed}$)
estimated using {\bf Model 1} is much closer to the expected NS radius
compared to that estimated from {\bf Model 2} (see Table 2 and Table 3).
\begin{figure*}
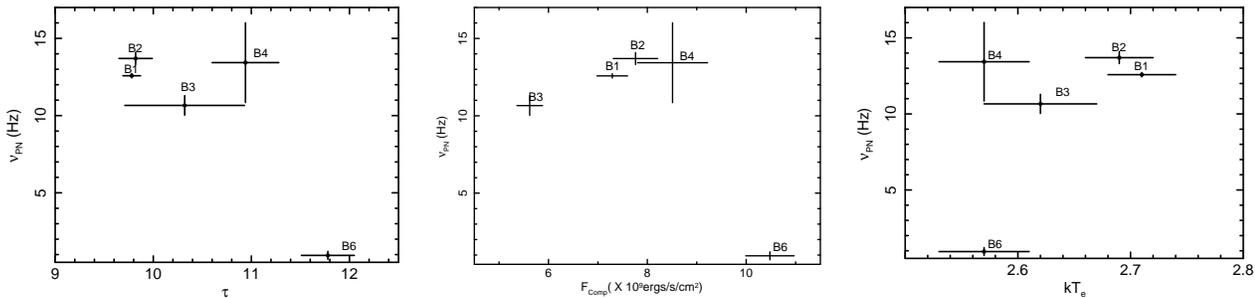

%\centering
%\begin{tabular}{@{}cc@{}}
\hspace{-.4in}
\includegraphics[width=0.22\textwidth,angle=-90]{cc-tau-pn-fq-rev.eps} %&
\includegraphics[width=0.22\textwidth, angle=-90]{cc-fcomp-pn-rev.eps} % \\
\includegraphics[width=0.22\textwidth,angle=-90]{cc-kte-pn-fq-rev.eps}
%\hspace{-1in}
%\end{tabular}
%\end{figure*}
%\begin{figure*}
%\includegraphics[width=0.35\textwidth,angle=-90]{cc-kte-pn-fq.eps}
\caption{Correlation between the spectral parameters and the best
fit PN frequency. Left plot (a) shows correlation between the optical
depth $\tau$ and $\nu_{PN}$. Centre plot (b) shows correlation between
$\nu_{PN}$ and the Comptonization flux, where as right plot (c) shows
correlation between kTe and $\nu_{PN}$. See text for details.}
\end{figure*}

In all seven parts of the HID (marked in Figure 2), we find
the presence of a power-law tail in the energy spectra. During our
observations, the source was in the soft spectral state. The source
has also shown a hard tail during the BeppoSAX observations
\citep{Piraino07,Piraino16} and Suzaku observations. During these
observations, the source was also in the soft state (see Piraino et
al. 2007; Lin et al 2010). A power-law tail has been previously
reported in atoll sources 4U 1728-34 \citep{Tarana11} and 4U 1636-536
\citep{Fiocchi06} and Z-sources GX 17+2 \citep{disalvo00}, Sco X-1
\citep{damico1}, GX 349+2 \citep{disalvo01} {\bf and} Cyg X-2 \citep{disalvo02}.

The contribution of {\bf the} power-law component to the total flux is around
$4-30$\% (for {\bf Model 1}). The flux does not show any clear correlation
with the position on the {\it banana pattern}. The flux was highest
(30\%) at the B3. There are many competing models to explain {\bf the} origin
of the non-thermal component. Such a hard tail can be produced
in either, hybrid thermal/non-thermal population of electrons in the
corona \citep{Poutanen98} or Comptonization in bulk motion of matter
close to the compact object \citep{Ebisawa96, Titarchuk98}. However,
\citet{Farinelli07} suggested that at a high accretion rate,
radiation pressure due to emission from the neutron star surface
can slow down the bulk motion of matter causing quenching of bulk
Comptonization.  Other possible mechanisms which have been proposed to
explain the non-thermal tails are, Comptonization of the seed
photons in a relativistic jet \citep{disalvo00} or synchrotron
emission from such a jet.

A broad iron line ($FWHM \sim$ 2 keV) is present in all the
spectra. Both {\it diskline} and a broad {\it Gaussian} line models provide equally good description of the iron line feature. This suggests that probably, smearing of the feature in a relativistic accretion disc account for the broadening.
An alternative mechanism, broadening by thermal Comptonization of
the line emission in a corona surrounding the disc can also explain the
large width (see \citealt{disalvo05}). In this case, line width is given
by \citep{kallman89, brandt94},
\begin{equation}
\sigma_{Fe} = 0.02E_{Fe} \tau (1+0.78 kT_e).
\end{equation}
In the present case, average temperature of the corona is $\sim$
2.7 keV and {\bf the} line energy is $\sim$ 6.5 keV. The line width is in the
range of $1.10-1.25$ keV (for {\bf Model 1}). Hence, the
Fe line is broadened in the corona having optical depth $\sim $
$2.7-3.2$. Therefore, most probably {\bf the} line is produced in the external
parts of the corona. Since in the present case {\bf the} optical depth is
very high ($\tau \sim$ $10-14$), the Fe-line produced at the inner
part of the corona will be completely smeared.

Correlated spectral and fast timing properties of this source have been
studied in detail by \cite{olive03} using the RXTE-PCA data. In the
soft state, they found low frequency noise (LFN) component centered at
$\sim$ $35-45$ Hz. VLFN was detected in one of the observations in the
soft state. We detect a broad feature  at $1-13$ Hz (PN) in the B1, B2,
B3 and B4 and B6 state. The integrated rms of this feature varies between
$1.87-7.1$\%, which is lower compared to that of the LFN. The
percentage rms of the LFN in the high soft state varies between
$5-12.5$\%. The PN frequencies (B1-B4) show correlation with the
total Comptonization flux in the $0.1-100$ keV band. \cite{olive03}
found that the LFN frequency is anti-correlated with {\bf the} hard flux
in the $20-200$ keV band. However, we find that at B6 though
Comptonization is strong frequency is low. We choose integrated
Comptonized flux instead of $20-200$ keV flux because it is a better
measure of strength of the corona. If this feature is produced by
oscillations in the corona, then it is expected that as the corona becomes
compact and cool, characteristic frequency of the oscillation may also
increase. But we find that the observed correlation between the
spectral and temporal properties is not consistent with this expected
behaviour. However, detection of the PN in the $3-20$ keV
band where the Comptonized emission is dominant suggests that the
PN is linked with the corona. It is also possible that the
PN is produced in the disc and modified in the corona as suggested
by \cite{swank01} to explain the low-frequency QPOs in the blackhole
candidates. A correlation between the disc flux and the
PN parameters will shed more light on this.
% The production of QPOs and noise features in the oscillating corona
%has been discussed by \cite{cabanac10}

\section{Conclusion}

In this work, we present a detailed spectral and timing analysis of the
source 4U 1705-44 using {\it LAXPC/AstroSat} data. The source traced a
{\it banana pattern} in the HID during the LAXPC observations. We fit
the spectra with three different approaches. We suggest that the
model consisting of Comptonized emission from the corona ({\bf Model
1}) is {\bf a} better description of the spectra in the {\it banana state}. The
spectra show presence of a variable hard tail and a broad iron line
centered at $6.4-6.7$ keV. Our analysis suggests that the broad
iron line is most probably produced at the outer part of the corona with
{\bf an} optical depth in the range of $2.7-3.2$. We also investigate the evolution of the spectral
parameters along the HID. We find that  {\bf the} optical depth of the corona
increases and the electron temperature decreases as the source moves
along the {\it banana branch}. Non-thermal component does not show
any clear correlation with the position on the HID. In the PDS, we find
a broad PN component (with {\bf a} quality factor Q $\sim$ $0.48-1.65$) centered
at $1-13$ Hz. The PN is detected in the energy band where Comptonization
is the main process, which suggests that {\bf the} origin of the PN is linked with
the corona. However, correlation between the disc/thermal flux and
the PN frequency is required in order to understand the mechanisms
producing the broad PN in the power spectra of this source.

\section*{Acknowledgements}

We thank the reviewer for providing us a set of useful comments and suggestions to improve the quality of the manuscript. This research has made use of the data obtained through GT phase of {\it AstroSat} observation. Authors thank GD, SAG; DD, PDMSA and Director, ISAC for encouragement and continuous support to
carry out this research.

\end{document}